\documentclass[twocolumn,a4paper,superscriptaddress,preprint,10pt,longbibliography]{revtex4-1}
\usepackage{graphicx}
\usepackage{subfigure}
\usepackage{umoline}
\usepackage{subfigure}
\usepackage{latexsym}   
\usepackage{amsmath,amssymb,amsthm}

\setcitestyle{super}

\renewcommand{\figurename}{{\bf Figure}}

\usepackage{color}

\begin{document}

\title{Universality in the morphology and mechanics of coarsening amyloid fibril networks}

\author{\firstname{L.} G. \surname{Rizzi}} %
\affiliation{School of Chemistry, University of Leeds, LS2 9JT, Leeds, UK.}
\author{\firstname{D.} A. \surname{Head}} %
\affiliation{School of Computing, University of Leeds, LS2 9JT, Leeds, UK.}
\author{\firstname{S.} \surname{Auer}} %
\affiliation{School of Chemistry, University of Leeds, LS2 9JT, Leeds, UK.}

\date{\today}


\begin{abstract}
\noindent
\bf 
	Above a critical concentration a wide variety of peptides and proteins self-assemble into amyloid fibrils which entangle to form percolating networks called hydrogels.
	Such hydrogels have important applications as biomaterials and in nanotechnology, but their applicability often depends on their mechanical properties for which we currently have no predictive capability.
	Here we use a peptide model to simulate the formation of amyloid fibril networks, and couple these to elastic network theory to determine their mechanical properties. 
	The simulations reveal that the time-dependence of morphological quantities characterizing the network length scales can be collapsed onto master curves by using a time scaling function that depends on the interaction parameter between the peptides. 
	The same scaling function is used to unveil a universal, non-monotonic dependence of the shear modulus with time.
	The obtained insight into the structure-function relationship between the peptide building blocks, network morphology and network mechanical properties can aid experimentalists to design amyloid fibril networks with tailored mechanical properties.
\end{abstract}


\maketitle

\noindent
	The formation of amyloid fibril networks proceeds in two stages. 
	Firstly, peptides or proteins assemble into amyloid fibrils that share a common cross-$\beta$ structure formed of intertwined layers of $\beta$-sheets extending in the direction parallel to the fibril axis~\cite{adamcik2012macromolecules,fitzpatrick2013pnas}.
	The lengthening and thickening of the fibrils can be attributed to strongly directional backbone hydrogen bonding and weaker side-chain interactions, respectively.
	At later times, the amyloid fibrils can entangle into a percolating network~\cite{schleeger2013polymer} with a morphology characterised by various lengths including the distance between crosslinks.
	Although amyloid fibrils are often associated with devastating diseases including Alzheimer's and Parkinson's disease~\cite{chiti2006annrevbiochem}, amyloid networks are emerging as an important class of material with applications in biosensing, nanoelectronics, tissue engineering and drug delivery~\cite{cherny2008angew,bowerman2012biopol,reynolds2014biomacromol}.
	However, many of these applications depend strongly on the network’s mechanical properties, for which we currently have no predictive structure-function relation between the bulk network stiffness and the properties of individual proteins without making assumptions regarding the fibril morphology, cross-linker dynamics, and the local deformation regime.

	Theoretical and experimental studies characterizing the mechanical properties of single, isolated amyloid fibrils show that their elastic modulus and bending stiffness are comparable to semiflexible filaments like actin, keratin, collagen and spider silk~\cite{knowles2006pnas,keten2010natmat,knowles2011nature,volpatti2013polymerphys}. 
	Recent rheology experiments demonstrate that the shear modulus of amyloid fibril networks can be altered by changes in the amino acid sequence of the peptides~\cite{tang2011langmuir,adler2014pccp} and the ionic strength of the solution~\cite{ozbas2004macromol,riley2009biotech,bolisetty2012biomacromolecules,saiani2013farad}, indicating that design principles for such networks might exist.
	The viscoelastic response of semiflexible polymer networks can be immediately related to the properties of individual filaments under the assumption of affinity, {\em i.e.} that the microscopic deformation field follows the applied macroscopic strain~\cite{pritchard2014accpt,broedersz2010prl,wolff2010newjphys,basu2011macromol,wen2012softmat}. Relaxing this assumption has thus far only been possible for the zero-frequency response of athermal networks, formally corresponding to the elastic plateau in rheological measurements where the crosslinks can be regarded as fixed~\cite{head2003prl,head2003pre,heussinger2006prl,broedersz2011np}.
	Importantly, all of the mentioned theoretical approaches assume that the filaments are identical, and thus have the same bending rigidity, or equivalently persistence length, at every point in the network.
	This assumption is a good approximation for filaments such as actin, but is not valid for amyloid fibril networks, where the fibril thickness varies throughout the network~\cite{kolsofszki2008progcolloid,xue2009protengdesig,morris2013natcomm}.
	Furthermore, these previous studies considered procedurally-defined, static networks; the time evolution of the network morphology was not considered.
	The objective of this study is to model the formation of amyloid fibril networks to determine their mechanical properties from elastic network theory without making the aforementioned assumptions.

\begin{figure*}[!t]
\centering
\includegraphics[width=0.9\textwidth]{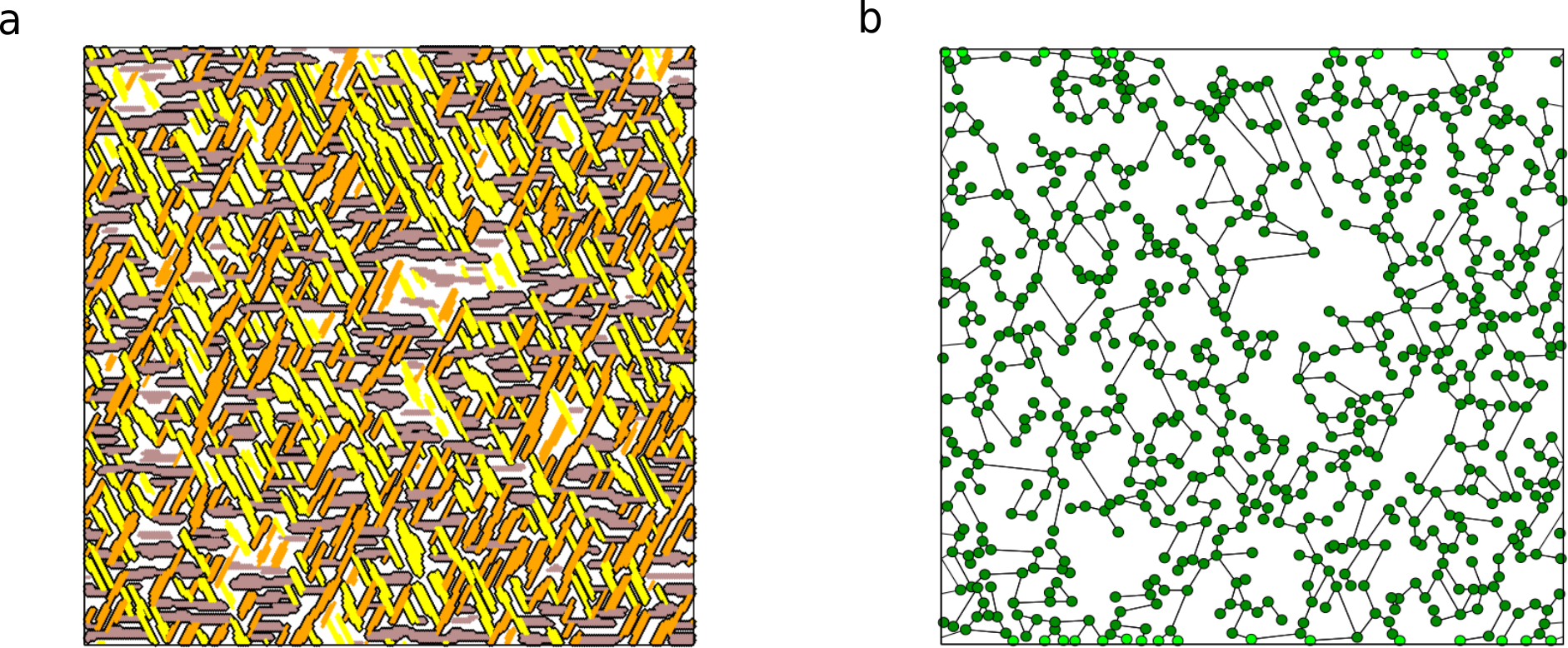}
\caption{{\bf Amyloid fibril network.}
{\bf a,} Amyloid fibril network obtained for $\xi=10$ with $N=31130$ peptides on a 2D lattice with linear size $L=256$ at time $t=10^5$~MC steps.
Fibrils that are part of the percolation network are shown with a black border.
The three possible orientations of the fibrils on the triangular lattice are distinguished by three different colours (yellow, brown, and orange). 
{\bf b,} Corresponding elastic network representation where fibril crosslinks are represented by circles and the fibrils between crosslinks by line segments.
}
\label{mapping_configs}
\end{figure*}

\vspace{0.5cm}

\noindent
{\bf Modelling amyloid fibril network formation}

\noindent
	Our simulations use a peptide model similar to those used in recent studies on the aggregation of amyloid fibrils~\cite{muthukumar2009jcp,cabriolu2012jcp,kashchiev2013jacs,irback2013prl}, where the peptides in their virtually fully extended ($\beta$-strand) conformation are described as hexagons positioned on a two-dimensional (2D) lattice with triangular symmetry (Supplementary Fig.~\ref{model}a).
	The use of a triangular lattice ensures that the mechanical properties of the fibril network are isotropic at large length scales~\cite{landauelasticity} as well as enabling the correct stacking of $\beta$-sheets in fibrils~\cite{hall2005jbiolchm} (Supplementary Fig.~\ref{model}b).
	The model shares similarities with the anisotropic Potts model, where only nearest-neighbor interactions along the three symmetry axes of the triangular lattice are considered.
	The hexagon has two opposing strong bonding sides that allow the formation of directional backbone hydrogen bonds. 
	This strong bond energy $E$ is denoted by the dimensionless parameter $\psi = E/k_B T$, where $k_B$ is Boltzmann's constant and $T$ the temperature.
	The remaining four sides of the hexagon allow the formation of weaker bonds modelling the side-chain interactions between the peptides.
	This bond energy $E_{h}$ is denoted $\psi_h = E_h/k_B T$.
	The $\beta$-strands can self-assembly laterally into $\beta$-sheets by forming strong hydrogen bonds that are parallel to the fibril lengthening axis, and the fibril can thicken due to the formation of weaker bonds between side-chains (Supplementary Fig.~\ref{model}b).
	Here we fix $\psi_h =1$ while varying $\psi$, to study effect of the ratio $\xi=\psi/\psi_h$ of the hydrogen bond energy to the side-chain bond energy.
	A value of $\xi=14$ has been estimated for the A$\beta_{40}$ peptide~\cite{kashchiev2013jacs}.

	In order to model the formation of amyloid fibril networks we perform Monte Carlo (MC) simulations similar to those described in ref.~\cite{muthukumar2009jcp}.
	We perform displacement moves of peptides to nearest neighbor lattice sites, and rotation moves so that the peptide can change its orientation.
	At the beginning of a simulation, $N$ peptides are randomly placed and oriented on a periodic 2D triangular lattice of linear size $L$. 
	In all our simulations we set $L=256$.
	As we only use physically plausible moves, there is a correspondence between the number of MC steps (MCs) and real time as suggested in refs.~\cite{muthukumar2009jcp,irback2013prl,linse2011molbios}.
	A typical configuration is shown in Fig.~\ref{mapping_configs}a.
	The fibril network obtained is very similar to those observed in experiments~\cite{karsai2007nanotech,li2009jphyschemB}.
	We use the Hoshen-Kopelman algorithm~\cite{hoshenkopelamn1976} to identify individual amyloid fibrils, which enable us to characterized them by their thickness $i$ and length $m$ (see Supplementary Fig.~\ref{model}b).
	The identification of fibrils also allows us to obtain a elastic network representation of the system as shown in Fig.~\ref{mapping_configs}b.
	The crosslink positions are taken to be the geometric centres of the overlap region between different fibrils (see Supplementary Fig.~\ref{crosslinkpos}), with the distance between two connected crosslinks defined as the crosslink length $l$.
	Note that not all fibrils identified in the system are part of the percolating network (Fig.~\ref{mapping_configs}a).

\begin{figure*}[!t]
\centering
\includegraphics[width=0.9\textwidth]{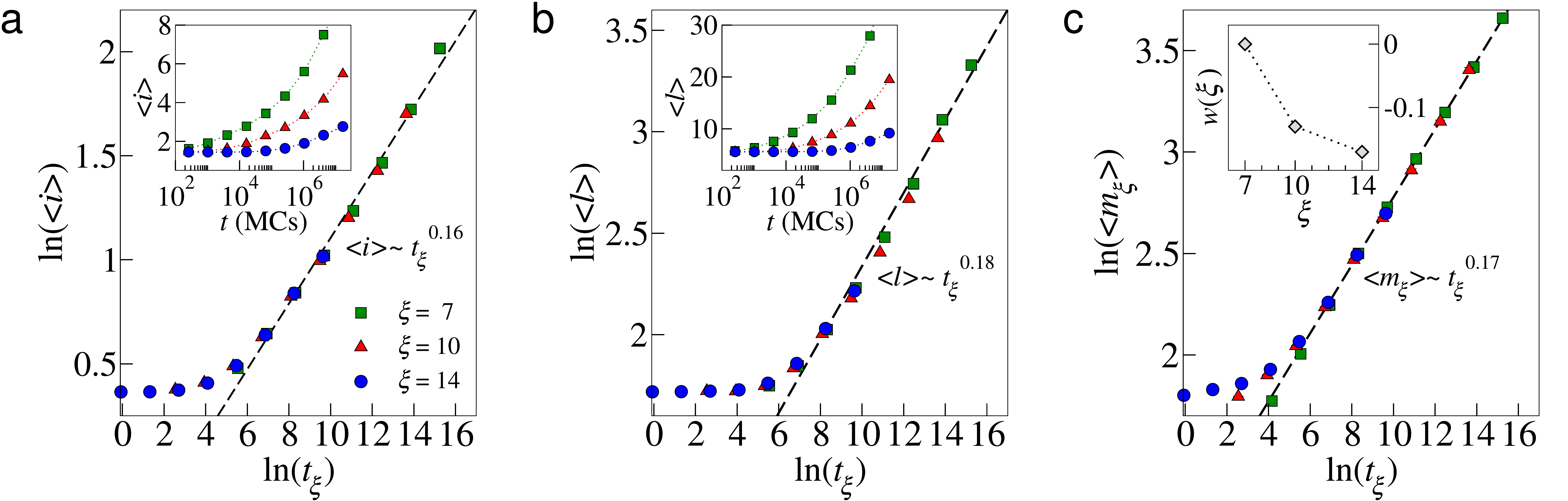}
\caption{{\bf Morphology of coarsening amyloid fibril networks.}
{\bf a,} Scaled and unscaled (inset) time dependence of the mean thickness $\langle i \rangle$.
{\bf b,} Scaled and unscaled (inset) time dependence of mean crosslink length $\langle l \rangle $.
{\bf c,} Scaled time dependence of mean fibril length $\langle m_{\xi} \rangle=\langle m \rangle e^{w(\xi)}$,  where the values of the function $w(\xi)$ used are shown in the inset.
The data is obtained from configurations for a coverage $\theta=0.5$ at times $t=4^n$ MCs, with $n=1,2,\dots,12$. Averages were obtained from $25$ independent simulations. Error bars are smaller than the symbols. Dashed lines are fits for $\ln(t_{\xi})>6$ as discussed in the text.}
\label{morph_scaling}
\end{figure*}


\vspace{0.5cm}

\noindent
{\bf Morphology of coarsening amyloid fibril networks}

\noindent
	The time evolution of the mean fibril thickness $\langle i \rangle$, the mean crosslink distance $\langle l \rangle$, and the mean fibril length $\langle m \rangle$ is shown in Fig.~\ref{morph_scaling} for anisotropy ratios $\xi=7$, $10$, and $14$, at a coverage $\theta=N/L^2=0.5$. 
	As can be seen in the insets to this figure, all lengths 
enter a coarsening regime where they monotonically increase, {\em i.e.} the fibrils become thicker and longer and the crosslink positions become further apart as the network evolves.
	Increasing the anisotropy ratio $\xi$ delays the onset of coarsening. 
	A similar effect, where the dynamics of the system is slowed down due to stronger effective interparticle interactions, has been observed in colloidal gels and glasses~\cite{lu2013annrev,hunter2012repprog}.
	As demonstrated in Fig.~\ref{morph_scaling}, the time dependence of each morphological quantity of the network obtained for different anisotropy ratios can be collapsed onto a single master curve by rescaling the time as $t_{\xi}=te^{-\Delta \xi}$, where $\Delta \xi = \xi - \xi_0$ with $\xi_{0}$ an arbitrary origin (here we take $\xi_{0}=7$).
	It is not only the mean of the morphological quantities that collapse, but also their distribution functions as shown in Supplementary Fig.~\ref{distributions}. 
	The coarsening exponents $\alpha_i = 0.16 \pm 0.02$, $\alpha_{l} = 0.18 \pm 0.02$, and $\alpha_m = 0.17 \pm 0.02$ of the morphological quantities can be 
	obtained by fitting the data points for times $\ln(t_{\xi}) > 6$ to 
	$\langle i \rangle \sim t_{\xi}^{\alpha_i}$, $\langle l \rangle \sim t_{\xi}^{\alpha_{l}}$, and $\langle m_{\xi} \rangle \sim  t_{\xi}^{\alpha_m}$, respectively.
	The mean fibril length was additionally scaled by an arbitrary function $w(\xi)$ (inset of Fig.~\ref{morph_scaling}c), giving $\langle m_{\xi} \rangle= \langle m \rangle e^{w(\xi)}$, but this does not affect the scaling with time.
	The measured exponents are consistent with a common value $\approx0.17$, demonstrating that all network lengths obey the same scaling during coarsening.
	Note that, if one identifies the fibril area $s\approx i\times m$ as a domain size, the growth law $s \sim t^{\alpha_i + \alpha_m}_{\xi}$ is consistent with the 1/3 exponent derived analytically for the isotropic case, i.e. the 2D spin-exchange Ising model~\cite{amar1988prb,mitchell2006prl}, despite the fibrils in our networks forming anisotropic shapes.

\vspace{0.5cm}

\noindent
{\bf Mechanics of amyloid fibril networks}

\noindent
	We now turn our attention to understanding how the structural changes in the network affects its mechanical properties.
	Starting from the network representation as in Fig.~\ref{mapping_configs}b, one can mimic an applied strain $\gamma$ by imposing horizontal displacements to the boundary crosslinks as shown in Supplementary Fig.~\ref{protocol}.
	The internal crosslinks then relax to new positions, causing an increase $\Delta E_{\text{elastic}}$ in the total elastic energy of the network.
	For the linear response ($\gamma \ll 1$), one can use $\Delta E_{\text{elastic}}$ to extract the shear modulus \cite{broedersz2011np,landauelasticity},
\begin{eqnarray}
G = \frac{2 \Delta E_{\text{elastic}}}{\gamma^2 A} ,
\label{shearmodulus}
\end{eqnarray}
where $A$ is the total system area.
	The shear modulus $G$ thus depends on the displacement vectors $\vec{u}_{\nu}$ of all crosslinks, which contribute to $\Delta E_{\text{elastic}}$ as detailed in Methods.
	These displacement vectors $\vec{u}_{\nu}$ are obtained by performing high-dimensional numerical optimization to minimize $\Delta E_{\text{elastic}}$ with respect to each displacement degree of freedom~\cite{head2003pre}.
	For all measurements we consider $\gamma=0.02$ to avoid non-linear responses due to high strain values.
	Our method generalises a previous lattice-based model \cite{broedersz2011np} by permitting variations in fibril thickness and crosslink distances (see Methods).

\begin{figure*}[!t]
\centering
\includegraphics[width=0.9\textwidth]{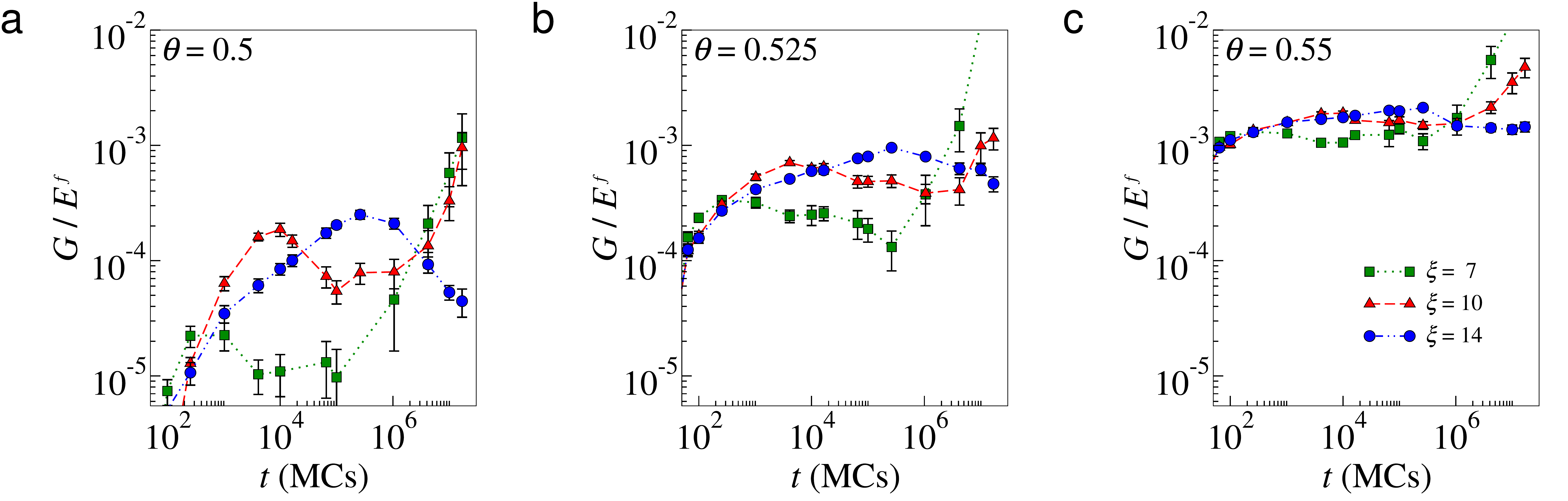}

\vspace{-0.2cm}

\caption{{\bf Time-dependence of the shear modulus.}
Panels {\bf a, b, c} corresponds to results obtained for networks with coverages $\theta=0.5$, $0.525$, and $0.55$, respectively.
Averages and errors bars were obtained from 25 independent simulations.
Additional points at times $t=10^n$ ($n \in \mathbb{N}$) are also included.}
\label{shearG_dif_theta0}
\end{figure*}

	In Fig. \ref{shearG_dif_theta0} we present results for the time-dependent behaviour of the normalized shear modulus $G/E^f$, where $E^f$ is the Young's modulus of a fibril.
	It is evident that small changes in both the anisotropy $\xi$ and the coverage $\theta$ lead to large changes in $G$ spanning orders of magnitude, with an overall trend for higher values of $G/E^f$ with increasing $\theta$. 
	Moreover, the variation is non-monotonic in time, in contrast to the monotonic coarsening of the morphological quantities discussed above, and there are time periods when the thinner fibrils (formed for $\xi=14$) yield stronger networks than thicker fibrils.
	Most experiments on the formation of amyloid fibril networks are performed on a time scale of tens of minutes, where the shear modulus displays either an increasing or a constant behaviour~\cite{aggeli1997nature,ozbas2004macromol,greenfield2010langmuir,bolisetty2012biomacromolecules}.
	However, one set of longer experiments suggested a slight decrease in $G$ after hours~\cite{gosal2004biomacromolecules_p2}, and atomic force microscopy imaging has demonstrated significant changes to network morphology over a timescale of days~\cite{karsai2007nanotech,usov2013faraddisc}, for which mechanical properties are not usually measured. This suggests extending the data acquisition window may reveal a similar non-monotonicity to Fig.~\ref{shearG_dif_theta0}.
	In addition, the behaviour of $G/E^{f}$ resembles that of weakly-interacting colloidal aggregates which measurably weaken prior to visual collapse~\cite{kilfoil2009}.
	We hypothesise this weakening shares a common mechanism to that observed in our results, although the final collapse under gravity (with its associated step change in the bulk symmetry) presumably has a different origin.

\begin{figure}[!t]
\centering
\includegraphics[width=0.4\textwidth]{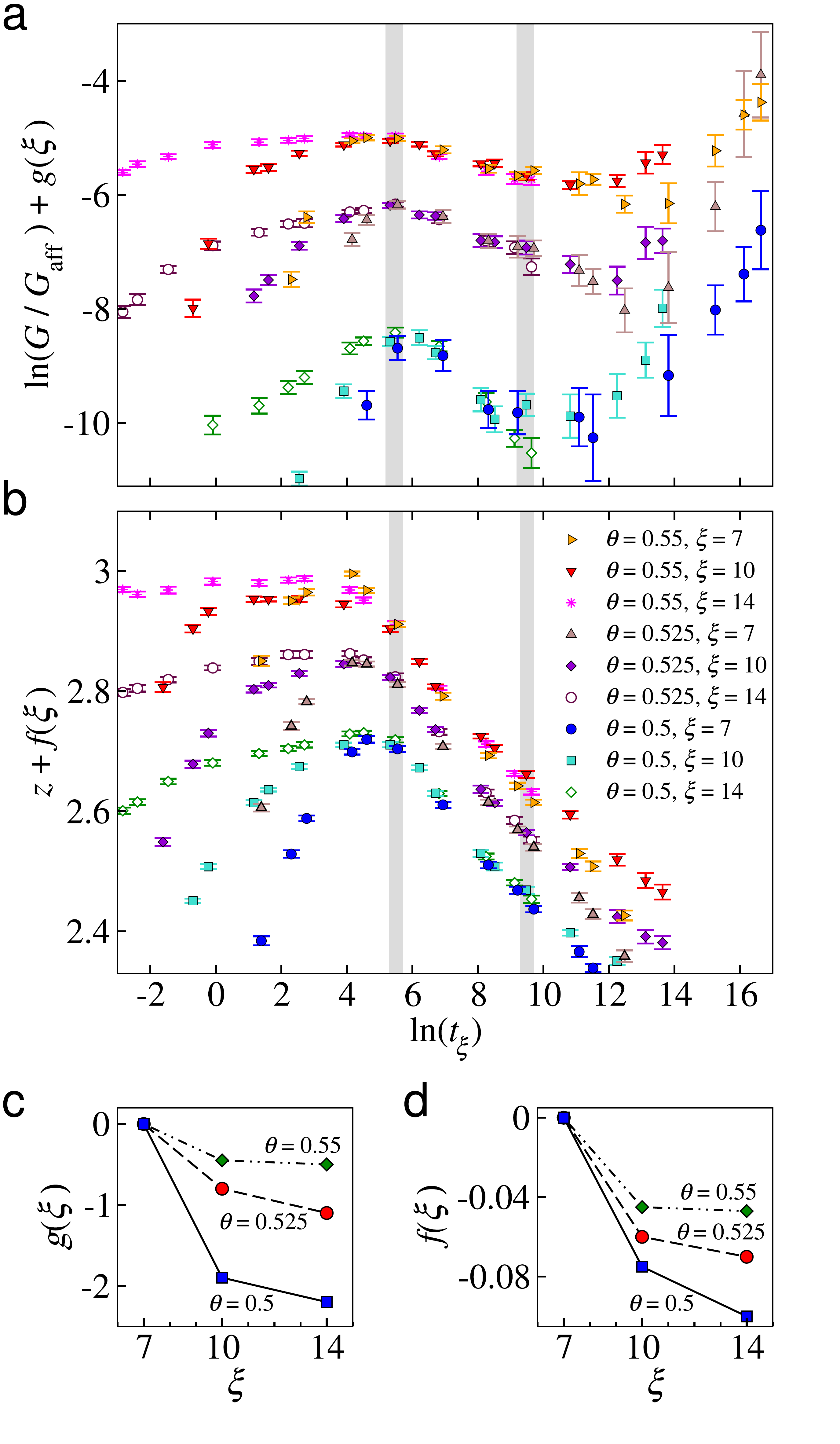}

\vspace{-0.5cm}

\caption{{\bf Time-scaling of the shear modulus. {\bf a,b}} Shear modulus ratio $G/G_{\text{aff}}$ and mean network connectivity $z$ against the rescaled time $t_{\xi}$, respectively.
Data collapse is obtained separately for each coverage $\theta$.
Vertical grey bars denote regions near the maximum and minimum of $G/E^f$.
{\bf c,d} Values of the scaling functions $g(\xi)$ and $f(\xi)$ for the shear modulus ratio and connectivity, respectively. }
\label{scaling_shearGGaff}
\end{figure}


	Insight into the mechanism underlying network weakening can be gained by measuring mechanical and morphological quantities simultaneously.
	We evaluated the shear modulus normalised to the affine prediction $G_{\text{aff}}$ (computed from Eq.~\ref{shearmodulus} for identical networks with affine displacements $\vec{u}_{\nu}^{\,\text{aff}}$) as in previous works~\cite{head2003pre,broedersz2011np}, and monitored the mean network connectivity $z$ given by the average value of the coordination numbers, $z_{\nu}$, from all internal crosslinks.
	As depicted in Fig.~\ref{scaling_shearGGaff}a, we found that $G/G_{\rm aff}$ for different anisotropies $\xi$ can be collapsed onto a series of master curves, with one such curve for each coverage~$\theta$. 
	The decreasing region of $G/G_{\text{aff}}$ coincides with the onset of a decrease in $z$ as indicated in the diagram, and indeed snapshots reveal a reduction in network connectivity over these times (see Supplementary Fig.~\ref{snapshots_SI}). 
	Thus, the mechanical weakening of the network is due to its increased sparsity.
	As with the fibril length $\langle m_{\xi}\rangle$ it was also necessary to scale the magnitude of $G/G_{\rm aff}$ by $\xi$-dependent factors $f(\xi)$ and $g(\xi)$ as shown in Figs.~\ref{scaling_shearGGaff}c and~\ref{scaling_shearGGaff}d, but this does not alter the times $t_{\xi}$ corresponding to the local minimum and maximum in~$G/G_{\rm aff}$.
	These factors mean that, at fixed rescaled times $t_{\xi}$, higher $\xi$ yields higher $G/G_{\text{aff}}$, and correspondingly higher values of $z$.
	For all values of $\theta$ and $\xi$, the values of $z$ are located below the central force threshold ($z_{\text{CF}}=4$ for 2D systems), and $G/G_{\rm aff}\ll1$ as evident in Fig.~\ref{scaling_shearGGaff}a.
	Thus all of our networks correspond to a non-affine deformation regime, and consistent with prior observations~\cite{broedersz2011np} most of the elastic energy takes the form of fibril bending, with $\Delta E_{\text{bending}}/\Delta E_{\text{elastic}}$ fluctuating around 0.9 (see Supplementary Fig.~\ref{energyratio}).

\vspace{0.5cm}

\noindent
{\bf Concluding remarks}

\noindent
	Identifying universal behaviours, such as in time-cure superposition curves~\cite{gosal2004biomacromolecules_p2,corrigan2009eurphysjE,corrigan2009langmuir}, can accelerate the development of novel materials by reducing the number of independent parameters that need to be assayed.
	Our simulations have revealed a simple time scaling function that depends on the anisotropy $\xi$ of interaction between peptides, which collapses data for both morphological and mechanical quantities.
	The proposed scaling function should benefit experimentalists in the design of amyloid-based materials, since it permits the extrapolation of the time-dependent mechanical response of the amyloid fibril networks from the behaviour of peptide systems with known interactions.
	Our findings indicate that features like the non-zero shear modulus for connectivities lower than $z_{\text{CF}}$ and the non-affine response, which are commonly overlooked in the modelling of hydrogels~\cite{basu2011macromol,wen2012softmat}, should be included in further descriptions of amyloid fibril networks.
	Finally, we note that our hybrid approach to measure the elastic moduli of a fiber network as it forms and grows represents a new direction for fiber network modelling that can be extended to other fibrous and porous materials in general, including inorganic materials such as colloidal gels~\cite{lu2013annrev}.

\vspace{0.5cm}
\noindent
{\bf Methods}

\vspace{0.3cm}

\noindent
{\small {\bf Elastic network model.}
Here we employ a modified version of a lattice-based elastic network model \cite{broedersz2011np} to evaluate $\Delta E_{\text{elastic}}$ as the sum of the changes in stretching and bending energies.	
The change in the stretching energy is defined as
\begin{eqnarray}
\Delta E_{\text{stretching}} = \frac{1}{2} \sum_{\nu < \mu} k_{\nu\mu} (\delta l_{\nu\mu} )^2 ~ ~ ~,
\label{stretchE}
\end{eqnarray}
where the sum is over all unique pairs of adjacent connected crosslinks $\nu$ and $\mu$, $k_{\nu\mu}$ is the spring constant of the fibril segment connecting them, and $\delta l_{\nu\mu}$ is the segment's extension due to the displacement vectors at each end, {\em i.e.} $\delta l_{\nu\mu}= \vec{u}_{\nu\mu} \cdot \hat{l}_{\nu\mu}$ with $\vec{u}_{\nu\mu} = \vec{u}_{\mu} - \vec{u}_{\nu}$  and $\hat{l}_{\nu \mu}= \vec{l}_{\nu \mu} / |\vec{l}_{\nu \mu}|$.
	The change in the bending energy is defined similarly,
\begin{eqnarray}
\Delta E_{\text{bending}} = \frac{1}{2} \sum_{ \nu\beta\mu }  \kappa_{\nu \beta \mu}  \frac{ \left(\delta \theta_{\nu\beta\mu} \right)^2 }{ \bar{l}_{\nu\beta\mu}}   ~ ~ ~,
\label{bendingE}
\end{eqnarray}
where the sum is over adjacent crosslinks along the same fibril, $\kappa_{\nu\beta\mu}$ is the bending rigidity of the corresponding fibril, $\bar{l}_{\nu\beta\mu} = (l_{\nu\beta} + l_{\beta\mu})/2$ is the mean crosslink length, and $\delta\theta_{\nu\beta\mu}$ denotes the change in angle between the two consecutive fibril segments $\nu\beta$ and $\beta\mu$.
	Note that the initial angle is generally not zero, but using the change in angle $\delta\theta_{\nu\beta\mu}$ means that the unstrained network has zero elastic energy, {\em i.e.} there are no pre-stresses.
	We estimate the coupling constants $k_{\nu\mu}$ and $\kappa_{\nu \beta \mu}$ by treating the fibrils as slender, defect-free elastic bodies with a uniform circular cross-section~\cite{landauelasticity}.
	Thus the spring constant and bending rigidity are written, respectively, as $k_{\nu\mu} = g_{\nu\mu} \pi R_{\nu \mu}^2 E^f / l_{\nu \mu}$ and $\kappa_{\nu\beta\mu} = g_{\nu\beta} \, g_{\beta\mu} \pi R_{\nu \beta}^4 E^f / 4$, where $R_{\nu \mu}$ is the radius of the cross-section of the fibril, $E^f$ its Young's modulus, and $g_{\nu\mu}=1$ for connected crosslinks or $0$ otherwise.
	The radius $R_{\nu \mu}$ is taken to be half of the fibril thickness $i_{\nu\mu}$ of the fibril segment linking the crosslink pair $\nu$ and $\mu$.
	We normalise both constants by $E^{f}$, giving $k^{\prime}_{\nu\mu} = k_{\nu\mu}/E^f = g_{\nu\mu} \pi i_{\nu \mu}^2 / 4 \, l_{\nu \mu}$ and $\kappa^{\prime}_{\nu\beta\mu} = \kappa_{\nu\beta\mu}/E^f= g_{\nu \beta} \, g_{\beta \mu} \pi i_{\nu \beta}^4 / 64$.}
	Note that although we have neglected the mechanical anisotropy of the fibrils, our key findings are expected to hold as long as the bending modulus scales with thickness as $\kappa_{\nu\beta\mu} \propto i_{\nu\beta}^4$.
	This is predicted by theoretical calculations for protein fibre bundles with bundling springs $\psi$ times weaker than fibre springs, separated by one monomer, for which $R^4$ scaling is expected~\cite{bathe2008bphysj} up to $R^2 \approx 100$, far in excess of our simulations.
	In addition, computational resources limited the network construction to 2D, but we employ 3D fibril properties so that our results can be experimentally realisable, for instance by confining the network to a Langmuir trough.
	Permitting out-of-plane displacements can significantly alter the measured moduli~\cite{muller2014prl}, but for actin networks the Langmuir trough was shown to suppress such modes~\cite{boatwright2011softmat} so we do not include these displacements in our calculations.

\small


\begin{thebibliography}{10}
\expandafter\ifx\csname url\endcsname\relax
  \def\url#1{\texttt{#1}}\fi
\expandafter\ifx\csname urlprefix\endcsname\relax\def\urlprefix{URL }\fi
\providecommand{\bibinfo}[2]{#2}
\providecommand{\eprint}[2][]{\url{#2}}

\bibitem{adamcik2012macromolecules}
\bibinfo{author}{Adamcik, J.} \& \bibinfo{author}{Mezzenga, R.}
\newblock \bibinfo{title}{Proteins fibrils from a polymer physics perspective}.
\newblock \emph{\bibinfo{journal}{Macromolecules}}
  \textbf{\bibinfo{volume}{45}}, \bibinfo{pages}{1137} (\bibinfo{year}{2012}).

\bibitem{fitzpatrick2013pnas}
\bibinfo{author}{Fitzpatrick, A. W.~P.} \emph{et~al.}
\newblock \bibinfo{title}{Atomic structure and hierarchical assembly of a
  cross-$\beta$ amyloid fibril}.
\newblock \emph{\bibinfo{journal}{P. Natl. Acad. Sci. USA}}
  \textbf{\bibinfo{volume}{110}}, \bibinfo{pages}{5468} (\bibinfo{year}{2013}).

\bibitem{schleeger2013polymer}
\bibinfo{author}{Schleegert, M.} \emph{et~al.}
\newblock \bibinfo{title}{Amyloids: From molecular structure to mechanical
  properties}.
\newblock \emph{\bibinfo{journal}{Polymer}} \textbf{\bibinfo{volume}{54}},
  \bibinfo{pages}{2473} (\bibinfo{year}{2013}).

\bibitem{chiti2006annrevbiochem}
\bibinfo{author}{Chiti, F.} \& \bibinfo{author}{Dobson, C.~M.}
\newblock \bibinfo{title}{Protein misfolding, functional amyloid, and human
  disease}.
\newblock \emph{\bibinfo{journal}{Annu. Rev. Biochem.}}
  \textbf{\bibinfo{volume}{75}}, \bibinfo{pages}{333} (\bibinfo{year}{2006}).

\bibitem{cherny2008angew}
\bibinfo{author}{Cherny, I.} \& \bibinfo{author}{Gazit, E.}
\newblock \bibinfo{title}{Amyloids: Not only pathological agents but also
  ordered nanomaterials}.
\newblock \emph{\bibinfo{journal}{Angew. Chem. Int. Ed.}}
  \textbf{\bibinfo{volume}{47}}, \bibinfo{pages}{4062} (\bibinfo{year}{2008}).

\bibitem{bowerman2012biopol}
\bibinfo{author}{Bowerman, C.~J.} \& \bibinfo{author}{Nilsson, B.~L.}
\newblock \bibinfo{title}{Self-assembly of amphipathic $\beta$-sheet peptides:
  Insights and applications}.
\newblock \emph{\bibinfo{journal}{Biopolymers}} \textbf{\bibinfo{volume}{98}},
  \bibinfo{pages}{169} (\bibinfo{year}{2012}).

\bibitem{reynolds2014biomacromol}
\bibinfo{author}{Reynolds, N.~P.}, \bibinfo{author}{Charnley, M.},
  \bibinfo{author}{Mezzenga, R.} \& \bibinfo{author}{Hartley, P.~G.}
\newblock \bibinfo{title}{Engineered lysozyme amyloid fibril networks support
  cellular growth and spreading}.
\newblock \emph{\bibinfo{journal}{Biomacromolecules}}
  \textbf{\bibinfo{volume}{15}}, \bibinfo{pages}{599} (\bibinfo{year}{2014}).

\bibitem{knowles2006pnas}
\bibinfo{author}{Smith, J.~F.}, \bibinfo{author}{Knowles, T. P.~J.},
  \bibinfo{author}{Dobson, C.~M.}, \bibinfo{author}{MacPhee, C.~E.} \&
  \bibinfo{author}{Welland, M.}
\newblock \bibinfo{title}{Characterization of the nanoscale properties of
  individual amyloid fibrils}.
\newblock \emph{\bibinfo{journal}{P. Natl. Acad. Sci. USA}}
  \textbf{\bibinfo{volume}{43}}, \bibinfo{pages}{15806} (\bibinfo{year}{2006}).


\bibitem{keten2010natmat}
\bibinfo{author}{Keten, S.}, \bibinfo{author}{Xu, B.},
  \bibinfo{author}{Ihle, B.} \&  \bibinfo{author}{Buehler, M.~J.}
\newblock \bibinfo{title}{Nanoconfinement controls stiffness, strength and mechanical toughness of beta-sheet crystals in silk}.
\newblock \emph{\bibinfo{journal}{Nat. Mater.}}
  \textbf{\bibinfo{volume}{9}}, \bibinfo{pages}{359} (\bibinfo{year}{2010}).


\bibitem{knowles2011nature}
\bibinfo{author}{Knowles, T. P.~J.} \& \bibinfo{author}{Buehler, M.~J.}
\newblock \bibinfo{title}{Nanomechanics of functional and pathological amyloid
  materials}.
\newblock \emph{\bibinfo{journal}{Nat. Nanotechnol.}}
  \textbf{\bibinfo{volume}{6}}, \bibinfo{pages}{469} (\bibinfo{year}{2011}).

\bibitem{volpatti2013polymerphys}
\bibinfo{author}{Volpatti, L.~R.} \& \bibinfo{author}{Knowles, T. P.~J.}
\newblock \bibinfo{title}{Polymer physics inspired approaches for the study of
  the mechanical properties}.
\newblock \emph{\bibinfo{journal}{J. Polymer Sci. B: Pol. Phys}}
  \textbf{\bibinfo{volume}{52}}, \bibinfo{pages}{281} (\bibinfo{year}{2014}).

\bibitem{tang2011langmuir}
\bibinfo{author}{Tang, C.}, \bibinfo{author}{Ulijn, R.} \&
  \bibinfo{author}{Saiani, A.}
\newblock \bibinfo{title}{Effect of glycine substitution on
  \uppercase{F}moc-diphenylalanine self-assembly and gelation properties}.
\newblock \emph{\bibinfo{journal}{Langmuir}} \textbf{\bibinfo{volume}{27}},
  \bibinfo{pages}{14438} (\bibinfo{year}{2011}).

\bibitem{adler2014pccp}
\bibinfo{author}{Adler, J.}, \bibinfo{author}{Scheidt, H.~A.},
  \bibinfo{author}{Kr\"uger, M.}, \bibinfo{author}{Thomas, L.} \&
  \bibinfo{author}{Huster, D.}
\newblock \bibinfo{title}{Local interactions influence the fibrillation
  kinetics, structure and dynamics of \uppercase{A}$\beta$(1-40) but leave the
  general fibril structure unchanged}.
\newblock \emph{\bibinfo{journal}{Phys. Chem. Chem. Phys.}}
  \textbf{\bibinfo{volume}{16}}, \bibinfo{pages}{7461} (\bibinfo{year}{2014}).

\bibitem{ozbas2004macromol}
\bibinfo{author}{Ozbas, B.~B.}, \bibinfo{author}{Rajagopal, K.~K.},
  \bibinfo{author}{Schneider, J. P.~J.} \& \bibinfo{author}{Pochan, D.~J.}
\newblock \bibinfo{title}{Salt-triggered peptide folding and consequent
  self-assembly into hydrogels with tunable modulus}.
\newblock \emph{\bibinfo{journal}{Macromolecules}}
  \textbf{\bibinfo{volume}{37}}, \bibinfo{pages}{7331} (\bibinfo{year}{2004}).

\bibitem{riley2009biotech}
\bibinfo{author}{Riley, J.~M.}, \bibinfo{author}{Aggeli, A.},
  \bibinfo{author}{Koopmans, R.~J.} \& \bibinfo{author}{McPherson, M.~J.}
\newblock \bibinfo{title}{Bioproduction and characterization of a
  p\uppercase{H} responsive self-assembling peptide}.
\newblock \emph{\bibinfo{journal}{Biotech. Bioeng.}}
  \textbf{\bibinfo{volume}{103}}, \bibinfo{pages}{241} (\bibinfo{year}{2009}).

\bibitem{bolisetty2012biomacromolecules}
\bibinfo{author}{Bolisetty, S.}, \bibinfo{author}{Harnau, L.},
  \bibinfo{author}{Jung, J.-M.} \& \bibinfo{author}{Mezzenga, R.}
\newblock \bibinfo{title}{Gelation, phase behaviour, and dynamics of
  $\beta$-lactoglobulin amyloid fibrils at varying concentrations and ionic
  strengths}.
\newblock \emph{\bibinfo{journal}{Biomacromolecules}}
  \textbf{\bibinfo{volume}{13}}, \bibinfo{pages}{3241} (\bibinfo{year}{2012}).

\bibitem{saiani2013farad}
\bibinfo{author}{Boothroyd, S.}, \bibinfo{author}{Miller, A.~F.} \&
  \bibinfo{author}{Saiani, A.}
\newblock \bibinfo{title}{From fibres to networks using self-assembling
  peptides}.
\newblock \emph{\bibinfo{journal}{Farad. Discuss.}}
  \textbf{\bibinfo{volume}{2013}}, \bibinfo{pages}{13} (\bibinfo{year}{2013}).

\bibitem{pritchard2014accpt}
\bibinfo{author}{Pritchard, R.}, \bibinfo{author}{Huang, Y. Y.~S.} \&
  \bibinfo{author}{Terentjev, E.~M.}
\newblock \bibinfo{title}{Mechanics of biological networks: From the cell
  cytoskeleton to connective tissue}.
\newblock \emph{\bibinfo{journal}{Soft Matter}} \textbf{\bibinfo{volume}{10}},
  \bibinfo{pages}{1864} (\bibinfo{year}{2014}).

\bibitem{broedersz2010prl}
\bibinfo{author}{Broedersz, C.~P.} \emph{et~al.}
\newblock \bibinfo{title}{Cross-link governed dynamics of biopolymers
  networks}.
\newblock \emph{\bibinfo{journal}{Phys. Rev. Lett.}}
  \textbf{\bibinfo{volume}{105}}, \bibinfo{pages}{238101}
  (\bibinfo{year}{2010}).

\bibitem{wolff2010newjphys}
\bibinfo{author}{Wolff, L.}, \bibinfo{author}{Fernandez, P.} \&
  \bibinfo{author}{Kroy, K.}
\newblock \bibinfo{title}{Inelastic mechanics of sticky biopolymer networks}.
\newblock \emph{\bibinfo{journal}{New J. Phys.}} \textbf{\bibinfo{volume}{12}},
  \bibinfo{pages}{053024} (\bibinfo{year}{2010}).

\bibitem{basu2011macromol}
\bibinfo{author}{Basu, A.} \emph{et~al.}
\newblock \bibinfo{title}{Nonaffine displacements in flexible polymer
  networks}.
\newblock \emph{\bibinfo{journal}{Macromolecules}}
  \textbf{\bibinfo{volume}{44}}, \bibinfo{pages}{1671} (\bibinfo{year}{2011}).

\bibitem{wen2012softmat}
\bibinfo{author}{Wen, Q.}, \bibinfo{author}{Basu, A.}, \bibinfo{author}{Janmey,
  P.~A.} \& \bibinfo{author}{Yodhb, A.~G.}
\newblock \bibinfo{title}{Non-affine deformations in polymer hydrogels}.
\newblock \emph{\bibinfo{journal}{Soft Matter}} \textbf{\bibinfo{volume}{8}},
  \bibinfo{pages}{8039} (\bibinfo{year}{2012}).

\bibitem{head2003prl}
\bibinfo{author}{Head, D.~A.}, \bibinfo{author}{Levine, A.} \&
  \bibinfo{author}{MacKintosh, F.}
\newblock \bibinfo{title}{Deformation of cross-linked semiflexible polymer
  networks}.
\newblock \emph{\bibinfo{journal}{Phys. Rev. Lett.}}
  \textbf{\bibinfo{volume}{91}}, \bibinfo{pages}{108102}
  (\bibinfo{year}{2003}).

\bibitem{head2003pre}
\bibinfo{author}{Head, D.~A.}, \bibinfo{author}{Levine, A.} \&
  \bibinfo{author}{MacKintosh, F.}
\newblock \bibinfo{title}{Distinct regimes of elastic response and deformation
  modes of cross-linked cytoskeletal and semiflexible polymer networks}.
\newblock \emph{\bibinfo{journal}{Phys. Rev. E}} \textbf{\bibinfo{volume}{68}},
  \bibinfo{pages}{061907} (\bibinfo{year}{2003}).


\bibitem{heussinger2006prl}
\bibinfo{author}{Heussinger, C.} \& \bibinfo{author}{Frey, E.}
\newblock \bibinfo{title}{Floppy modes and nonaffine deformations in random fiber networks}.
\newblock \emph{\bibinfo{journal}{Phys. Rev. Lett.}} \textbf{\bibinfo{volume}{97}},
  \bibinfo{pages}{105501} (\bibinfo{year}{2006}).


\bibitem{broedersz2011np}
\bibinfo{author}{Broedersz, C.~P.}, \bibinfo{author}{Mao, X.},
  \bibinfo{author}{Lubensky, T.~C.} \& \bibinfo{author}{MacKintosh, F.~C.}
\newblock \bibinfo{title}{Criticality and isostaticity in fibre networks}.
\newblock \emph{\bibinfo{journal}{Nat. Phys.}} \textbf{\bibinfo{volume}{7}},
  \bibinfo{pages}{983} (\bibinfo{year}{2011}).

\bibitem{kolsofszki2008progcolloid}
\bibinfo{author}{Kolsofszki, M.}, \bibinfo{author}{Karsai, A.},
  \bibinfo{author}{So\'os, K.}, \bibinfo{author}{Penke, B.} \&
  \bibinfo{author}{Kellermayer, M. S.~Z.}
\newblock \bibinfo{title}{Thermally-induced effects in oriented network of
  amyloid $\beta$25-35 fibrils}.
\newblock \emph{\bibinfo{journal}{Prog. Colloid Polym. Sci.}}
  \textbf{\bibinfo{volume}{135}}, \bibinfo{pages}{169} (\bibinfo{year}{2008}).

\bibitem{xue2009protengdesig}
\bibinfo{author}{Xue, W.~F.}, \bibinfo{author}{Homans, S.~W.} \&
  \bibinfo{author}{Radford, S.~E.}
\newblock \bibinfo{title}{Amyloid fibril length distribution quantified by
  atomic force microscopy single-particle image analysis}.
\newblock \emph{\bibinfo{journal}{Protein Eng. Des. Sel.}}
  \textbf{\bibinfo{volume}{22}}, \bibinfo{pages}{489} (\bibinfo{year}{2009}).

\bibitem{morris2013natcomm}
\bibinfo{author}{Morris, R.~J.} \emph{et~al.}
\newblock \bibinfo{title}{Mechanistic and enviromental control of the
  prevalence and lifetime of amyloid oligomers}.
\newblock \emph{\bibinfo{journal}{Nat. Comm.}} \textbf{\bibinfo{volume}{4}},
  \bibinfo{pages}{1891} (\bibinfo{year}{2013}).

\bibitem{muthukumar2009jcp}
\bibinfo{author}{Zhang, J.} \& \bibinfo{author}{Muthukumar, M.}
\newblock \bibinfo{title}{Simulations of nucleation and elongation of amyloid
  fibrils}.
\newblock \emph{\bibinfo{journal}{J. Chem. Phys.}}
  \textbf{\bibinfo{volume}{130}}, \bibinfo{pages}{035102}
  (\bibinfo{year}{2009}).

\bibitem{cabriolu2012jcp}
\bibinfo{author}{Cabriolu, R.}, \bibinfo{author}{Kashchiev, D.} \&
  \bibinfo{author}{Auer, S.}
\newblock \bibinfo{title}{Breakdown of nucleation theory for crystals with
  strongly anisotropic interactions between molecule}.
\newblock \emph{\bibinfo{journal}{J. Chem. Phys.}}
  \textbf{\bibinfo{volume}{137}}, \bibinfo{pages}{204903}
  (\bibinfo{year}{2012}).

\bibitem{kashchiev2013jacs}
\bibinfo{author}{Kashchiev, D.}, \bibinfo{author}{Cabriolu, R.} \&
  \bibinfo{author}{Auer, S.}
\newblock \bibinfo{title}{Confounding the paradigm: Peculiarities of amyloid
  fibril nucleation}.
\newblock \emph{\bibinfo{journal}{J. Am. Chem. Soc.}}
  \textbf{\bibinfo{volume}{135}}, \bibinfo{pages}{1531} (\bibinfo{year}{2013}).

\bibitem{irback2013prl}
\bibinfo{author}{Irback, A.}, \bibinfo{author}{Jonsson, S.},
  \bibinfo{author}{Linnemann, N.}, \bibinfo{author}{Linse, B.} \&
  \bibinfo{author}{Wallin, S.}
\newblock \bibinfo{title}{Aggregate geometry in amyloid fibril nucleation}.
\newblock \emph{\bibinfo{journal}{Phys. Rev. Lett.}}
  \textbf{\bibinfo{volume}{110}}, \bibinfo{pages}{058101}
  (\bibinfo{year}{2013}).

\bibitem{landauelasticity}
\bibinfo{author}{Landau, L.~D.} \& \bibinfo{author}{Lifshitz, E.~M.}
\newblock \emph{\bibinfo{title}{Theory of \uppercase{E}lasticity}}
  (\bibinfo{publisher}{Butterworth-Heineman Oxford}, \bibinfo{year}{1986}).

\bibitem{hall2005jbiolchm}
\bibinfo{author}{Nguyen, H.~D.} \& \bibinfo{author}{Hall, C.~K.}
\newblock \bibinfo{title}{Kinetics of fibril formation by polyalanine
  peptides}.
\newblock \emph{\bibinfo{journal}{J. Biol. Chem.}}
  \textbf{\bibinfo{volume}{280}}, \bibinfo{pages}{9074} (\bibinfo{year}{2005}).

\bibitem{linse2011molbios}
\bibinfo{author}{Linse, B.} \& \bibinfo{author}{Linse, S.}
\newblock \bibinfo{title}{Monte \uppercase{C}arlo simulations of protein
  amyloid formation reveal origin of sigmoidal aggregation kinetics}.
\newblock \emph{\bibinfo{journal}{Mol. Biosyst.}} \textbf{\bibinfo{volume}{7}},
  \bibinfo{pages}{2296} (\bibinfo{year}{2011}).

\bibitem{karsai2007nanotech}
\bibinfo{author}{Karsai, A.} \emph{et~al.}
\newblock \bibinfo{title}{Potassium-dependent oriented growth of amyloid
  $\beta$25-35 fibrils on mica}.
\newblock \emph{\bibinfo{journal}{Nanotechnology}}
  \textbf{\bibinfo{volume}{18}}, \bibinfo{pages}{345102}
  (\bibinfo{year}{2007}).

\bibitem{li2009jphyschemB}
\bibinfo{author}{Li, H.} \emph{et~al.}
\newblock \bibinfo{title}{Peptide diffusion and self-assembly in ambient water
  nanofilm on mica surfaces}.
\newblock \emph{\bibinfo{journal}{J. Phys. Chem. B}}
  \textbf{\bibinfo{volume}{113}}, \bibinfo{pages}{8795} (\bibinfo{year}{2009}).

\bibitem{hoshenkopelamn1976}
\bibinfo{author}{Hoshen, J.} \& \bibinfo{author}{Kopelman, R.}
\newblock \bibinfo{title}{Percolation and cluster distribution. \uppercase{I}.
  \uppercase{C}luster multiple labeling technique and critical concentration
  algorithm}.
\newblock \emph{\bibinfo{journal}{Phys. Rev. B}} \textbf{\bibinfo{volume}{14}},
  \bibinfo{pages}{3438} (\bibinfo{year}{1976}).

\bibitem{lu2013annrev}
\bibinfo{author}{Lu, P.~J.} \& \bibinfo{author}{Weitz, D.~A.}
\newblock \bibinfo{title}{Colloidal particles: crystals, glasses, and gels}.
\newblock \emph{\bibinfo{journal}{Annu. Rev. Condens. Matter Phys.}}
  \textbf{\bibinfo{volume}{4}}, \bibinfo{pages}{217} (\bibinfo{year}{2013}).

\bibitem{hunter2012repprog}
\bibinfo{author}{Hunter, G.~L.} \& \bibinfo{author}{Weeks, E.~R.}
\newblock \bibinfo{title}{The physics of the colloidal glass transition}.
\newblock \emph{\bibinfo{journal}{Rep. Prog. Phys.}}
  \textbf{\bibinfo{volume}{75}}, \bibinfo{pages}{066501}
  (\bibinfo{year}{2012}).

\bibitem{amar1988prb}
\bibinfo{author}{Amar, J.~G.}, \bibinfo{author}{Sulivan, F.~E.} \&
  \bibinfo{author}{Mountain, R.~D.}
\newblock \bibinfo{title}{Monte \uppercase{C}arlo study of growth in the
  two-dimensional spin-exchange kinetic \uppercase{I}sing model}.
\newblock \emph{\bibinfo{journal}{Phys. Rev. B}} \textbf{\bibinfo{volume}{37}},
  \bibinfo{pages}{196} (\bibinfo{year}{1988}).

\bibitem{mitchell2006prl}
\bibinfo{author}{Mitchell, S.~J.} \& \bibinfo{author}{Landau, D.~P.}
\newblock \bibinfo{title}{Phase separation in a compressible 2\uppercase{D}
  \uppercase{I}sing model}.
\newblock \emph{\bibinfo{journal}{Phys. Rev. Lett.}}
  \textbf{\bibinfo{volume}{97}}, \bibinfo{pages}{025701}
  (\bibinfo{year}{2006}).

\bibitem{aggeli1997nature}
\bibinfo{author}{Aggeli, A.} \emph{et~al.}
\newblock \bibinfo{title}{Responsive gels formed by the spontaneous
  self-assembly of peptides into polymeric $\beta$-sheet tapes}.
\newblock \emph{\bibinfo{journal}{Nature}}  (\bibinfo{year}{1997}).

\bibitem{greenfield2010langmuir}
\bibinfo{author}{Greenfield, M.~A.}, \bibinfo{author}{Hoffman, J.~R.},
  \bibinfo{author}{de~la Cruz, M.~O.} \& \bibinfo{author}{Stupp, S.~I.}
\newblock \bibinfo{title}{Tunable mechanics of peptide nanofiber gels}.
\newblock \emph{\bibinfo{journal}{Langmuir}} \textbf{\bibinfo{volume}{26}},
  \bibinfo{pages}{3641} (\bibinfo{year}{2010}).

\bibitem{gosal2004biomacromolecules_p2}
\bibinfo{author}{Gosal, W.~S.}, \bibinfo{author}{Clark, A.~H.} \&
  \bibinfo{author}{Ross-Murphy, S.~B.}
\newblock \bibinfo{title}{Fibrillar $\beta$-lactoglobulin gels: part 2. dynamic
  mechanical characterization of heat-set systems}.
\newblock \emph{\bibinfo{journal}{Biomacromolecules}}
  \textbf{\bibinfo{volume}{5}}, \bibinfo{pages}{2420} (\bibinfo{year}{2004}).

\bibitem{usov2013faraddisc}
\bibinfo{author}{Usov, I.}, \bibinfo{author}{Adamcik, J.} \&
  \bibinfo{author}{Mezzenga, R.}
\newblock \bibinfo{title}{Polymorphism in bovine serum albumin fibrils:
  morphology and statistical analysis}.
\newblock \emph{\bibinfo{journal}{Farad. Discuss.}}
  \textbf{\bibinfo{volume}{166}}, \bibinfo{pages}{151} (\bibinfo{year}{2013}).

\bibitem{kilfoil2009}
\bibinfo{author}{Kamp, S.~W.} \& \bibinfo{author}{Kilfoil, M.~L.}
\newblock \bibinfo{title}{Universal behavior in the mechanical properties of
  weakly aggregated colloidal particles}.
\newblock \emph{\bibinfo{journal}{Soft Matter}} \textbf{\bibinfo{volume}{5}},
  \bibinfo{pages}{2438} (\bibinfo{year}{2009}).

\bibitem{corrigan2009eurphysjE}
\bibinfo{author}{Corrigan, A.~M.} \& \bibinfo{author}{Donald, A.~M.}
\newblock \bibinfo{title}{Particle tracking microrheology of gel-forming
  amyloid fibril networks}.
\newblock \emph{\bibinfo{journal}{Eur. Phys. J. E}}
  \textbf{\bibinfo{volume}{28}}, \bibinfo{pages}{457} (\bibinfo{year}{2009}).

\bibitem{corrigan2009langmuir}
\bibinfo{author}{Corrigan, A.~M.} \& \bibinfo{author}{Donald, A.~M.}
\newblock \bibinfo{title}{Passive microrheology of solvent-induced fibrillar
  protein networks}.
\newblock \emph{\bibinfo{journal}{Langmuir}} \textbf{\bibinfo{volume}{25}},
  \bibinfo{pages}{8599} (\bibinfo{year}{2009}).



\bibitem{bathe2008bphysj}
\bibinfo{author}{Bathe, M.},  \bibinfo{author}{Heussinger, C.},  \bibinfo{author}{Claessens, M.~M.~A.~E.},  \bibinfo{author}{Bausch, A.~R.} \& \bibinfo{author}{Frey, E.}
\newblock \bibinfo{title}{Cytoskeletal bundle mechanics}.
\newblock \emph{\bibinfo{journal}{Biophys. J.}} \textbf{\bibinfo{volume}{94}},
  \bibinfo{pages}{2955} (\bibinfo{year}{2008}).


\bibitem{muller2014prl}
\bibinfo{author}{M\"uller, P.} \& \bibinfo{author}{Kierfeld, J.}
\newblock \bibinfo{title}{Wrinkling of random and regular semiflexible polymer networks}.
\newblock \emph{\bibinfo{journal}{Phys. Rev. Lett.}} \textbf{\bibinfo{volume}{112}},
  \bibinfo{pages}{094303} (\bibinfo{year}{2014}).

\bibitem{boatwright2011softmat}
\bibinfo{author}{Boatwright, T.},
\bibinfo{author}{Levine, A.~J.} \& \bibinfo{author}{Dennin, M.}
\newblock \bibinfo{title}{Mechanical reorganization of cross-linked \uppercase{F}-actin networks at the air-buffer interface}.
\newblock \emph{\bibinfo{journal}{Soft Matter}} \textbf{\bibinfo{volume}{7}},
  \bibinfo{pages}{7851} (\bibinfo{year}{2011}).

\end{thebibliography}

\vspace{0.5cm}

\noindent
{\bf Acknowledgements}\\
{\small
	L.G.R. acknowledges support from the Brazilian agency CNPq (Grant N$^{\text{o}}$ 245412/2012-3). 
	D.A.H. acknowledges support from the Biomedical Health Research Centre, University of Leeds, UK.}

\vspace{0.5cm}

\noindent
{\bf Author contributions}\\
{\small S.A., D.A.H., and L.G.R conceived and designed the experiments. L.G.R. performed the experiments. S.A., D.A.H., and L.G.R. analysed the data. All authors discussed the results, and wrote the manuscript.}

\vspace{0.5cm}

\noindent
{\bf Additional information}\\
{\small Supplementary information is attached in this arXiv version of the paper.
}

\vspace{0.5cm}

\noindent
{\bf Competing financial interests}\\
{\small The authors declare no competing financial interests.}




\pagebreak
\widetext

\renewcommand{\figurename}{{\bf Supplementary Figure}}

\noindent
{\bf Supplementary Information: Universality in the morphology and mechanics of coarsening amyloid fibril networks.}\\
L. G. Rizzi, D. A. Head, S. Auer

\setcounter{figure}{0}

\vspace{0.5cm}

\begin{figure}[h!]
\centering
\includegraphics[width=0.85\textwidth]{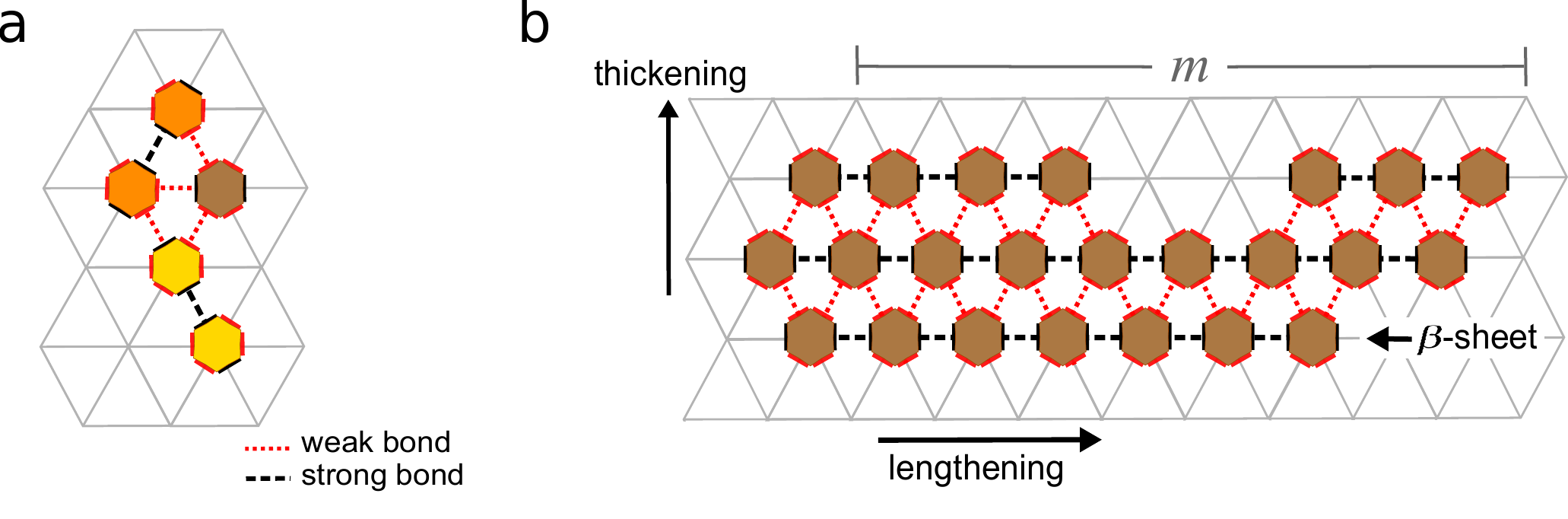}

\vspace{-0.2cm}
\caption{{\bf Peptide model.} {\bf a,} Interacting peptides in their $\beta$-strand conformation are represented as hexagons. 
Each hexagon has two opposing strong bonding sides (shown in black) that allow the formation of directional backbone hydrogen bonds (black dashed lines) and
four sides  (shown in red) that allow the formation of weaker bonds (red dotted lines) corresponding to the interactions between side-chains.
The three possible orientations of the hexagons on a triangular lattice are distinguished by the three different colours (yellow, brown, and orange).
{\bf b,} Amyloid fibril composed of three $\beta$-sheet layers. 
The hexagons self-assemble laterally into $\beta$-sheets by forming strong hydrogen bonds that are parallel to the fibril lengthening axis, and fibril thickening is due the formation of weaker bonds. The fibril length $m$ is taken to be the number of lattice sites between the peptides at the two fibril ends (in the direction of the fibril lengthening axis), while the fibril thickness is taken to be $i=n_p/m$, where $n_p$ is the total number of peptides in the fibril. For the shown fibril $m=9$ and $i= 23 /9 = 2.6$.}
\label{model}
\end{figure}
                                              
\vspace{1.5cm}
                                                                                       
\begin{figure}[h!]
\centering
\includegraphics[width=0.75\textwidth]{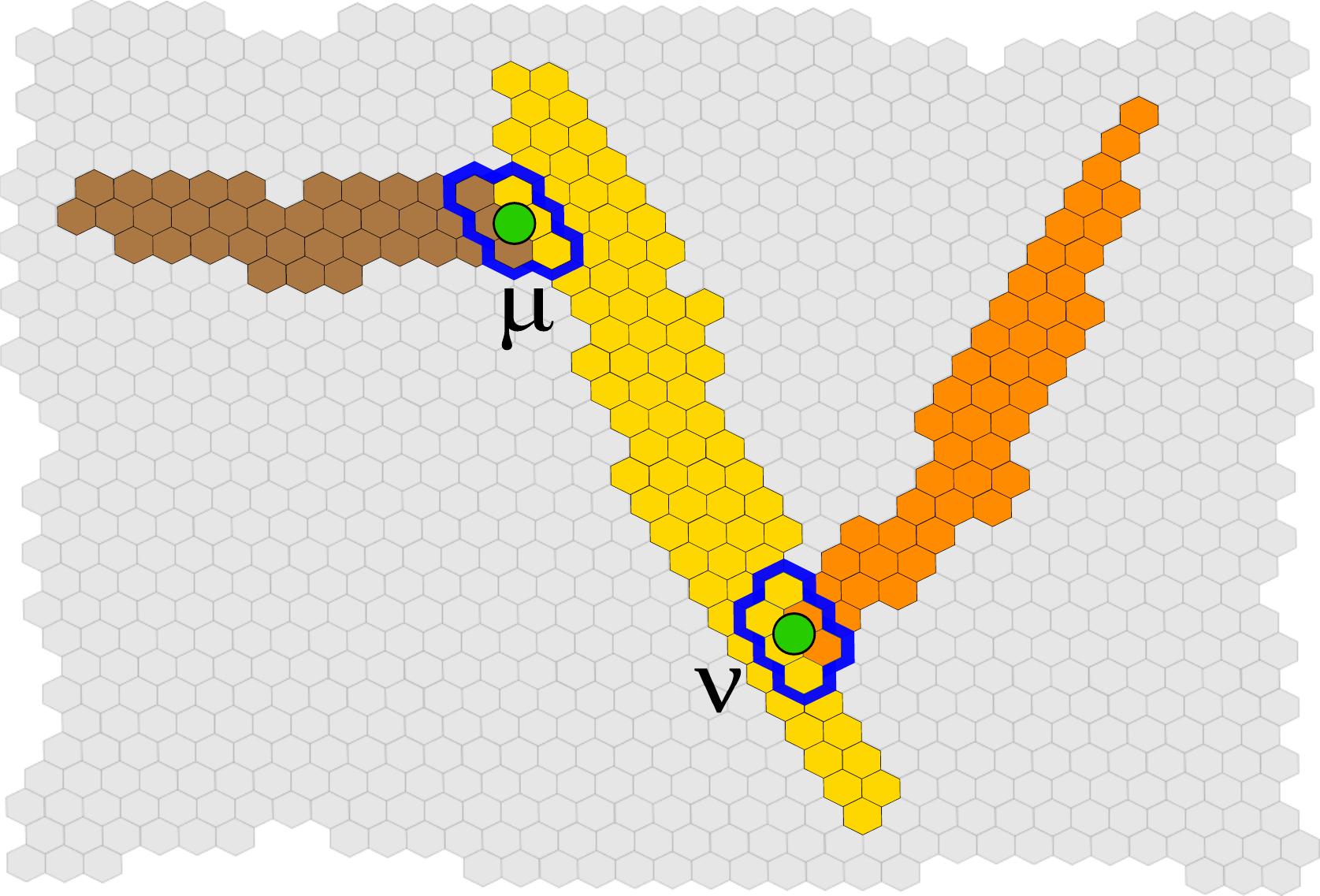}
\caption{{\bf Network crosslinks.}
The crosslink positions $\nu$ and $\mu$ (green circles) of three amyloid fibrils
are taken to be the geometric centre of the positions of the peptides in the crosslink region (highlighted with blue border). 
The crosslink length $l_{\nu\mu}$ is the distance between the the crosslinks $\nu$ and $\mu$, and the thickness $i_{\nu\mu}$
of the fibril segment linking the crosslink pair $\nu$ and $\mu$ is taken to be the thickness $i$ of the fibril connecting them.}
\label{crosslinkpos}
\end{figure}

\begin{figure}[h!]
\centering
\includegraphics[width=0.95\textwidth]{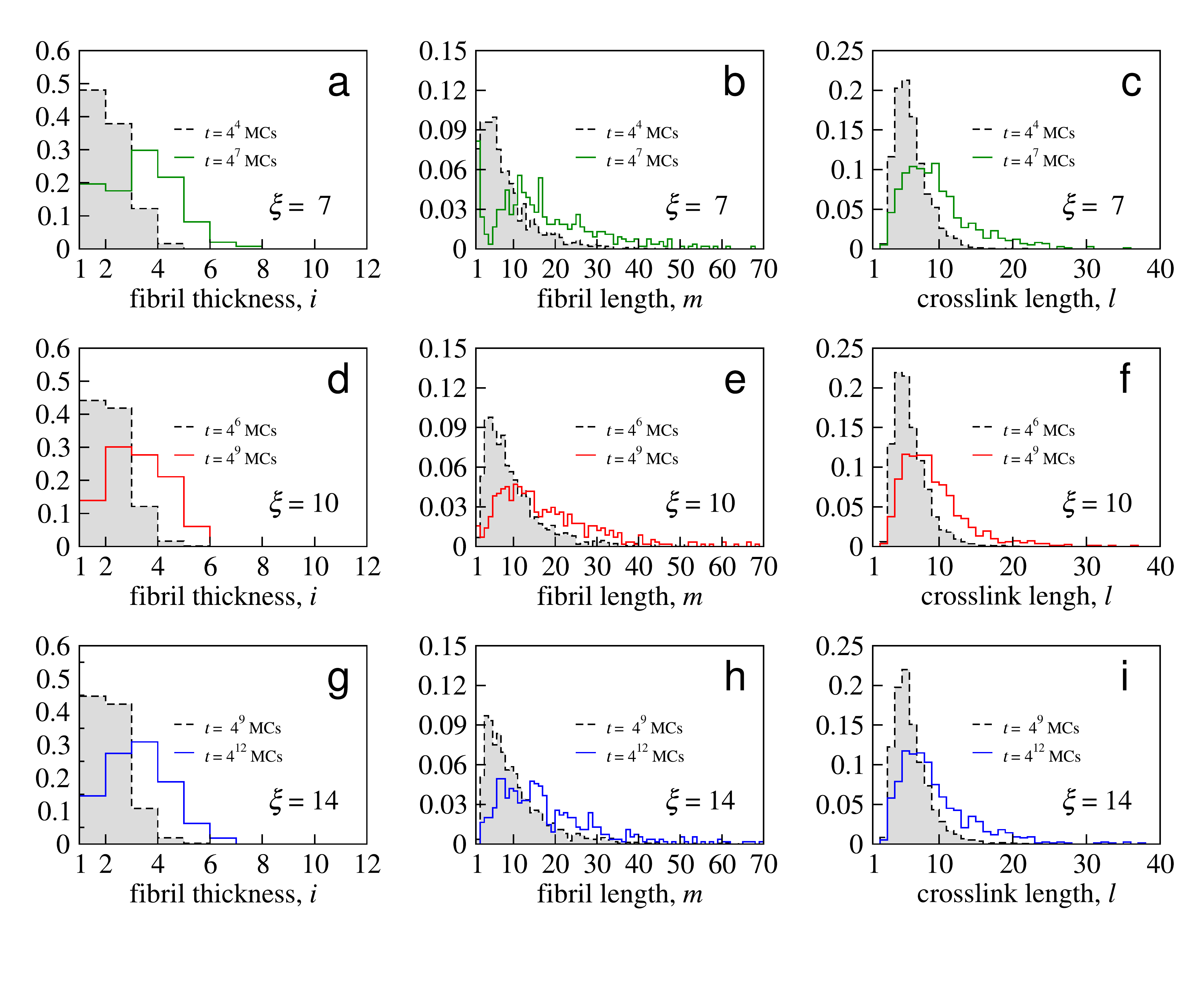}

\vspace{-0.5cm}

\caption{{\bf Distribution functions of the morphological quantities of the network.}
For each value of $\xi$ the histograms correspond to two representative configurations taken at the rescaled times $\ln(t_{\xi}) \sim 5.4$ and $\sim 9.4$ for a fixed coverage $\theta=0.5$ (snapshots of the corresponding configurations are displayed in Supplementary Fig. \ref{snapshots_SI} below).
Considering the values of anisotropy $\xi=7$, $10$, and $14$, the panels {\bf a, d,} and {\bf g} shows histograms for the fibril thickness, $i$, panels {\bf b, e,} and {\bf h} presents histograms for the fibril length, $m$, and panels {\bf c, f,} and {\bf i} displays histograms for the crosslink length, $l$.
Although the configurations were taken at different times $t$ and from systems with different values of $\xi$, the distributions are very similar.}
\label{distributions}
\end{figure}

\begin{figure}[h!]
\centering
\includegraphics[width=0.68\textwidth]{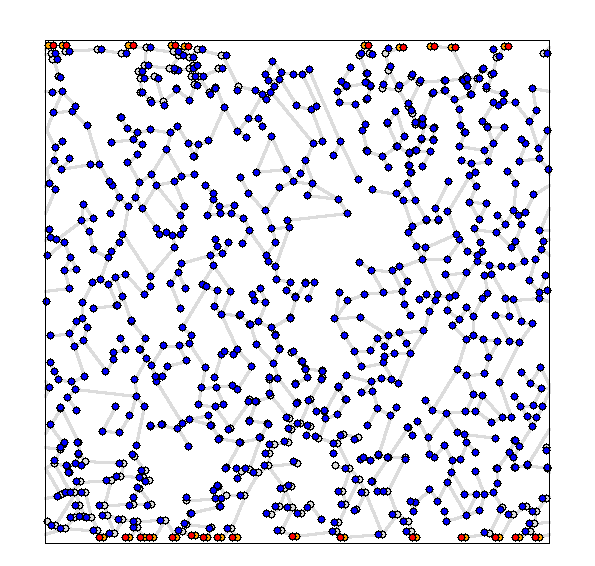}
\caption{{\bf Reference and sheared configurations of the amyloid fibril network.}
The elastic network representation of the system enable us to mimic an applied strain $\gamma$ by imposing a horizontal displacement to the boundary crosslinks (the orange and red circles correspond to the position of the reference and sheared crosslinks, respectively). The positions of the internal crosslinks of the sheared configuration (blue circles) are obtained by a high-dimensional numerical optimization algorithm to minimize the change in the elastic energy $\Delta E_{\text{elastic}}$ of the network, which allows us to determine the shear modulus $G$ of a given configuration using Eq. \ref{shearmodulus} of the main text.}
\label{protocol}
\end{figure}

\begin{figure}[h!]
\centering
\includegraphics[width=0.54\textwidth]{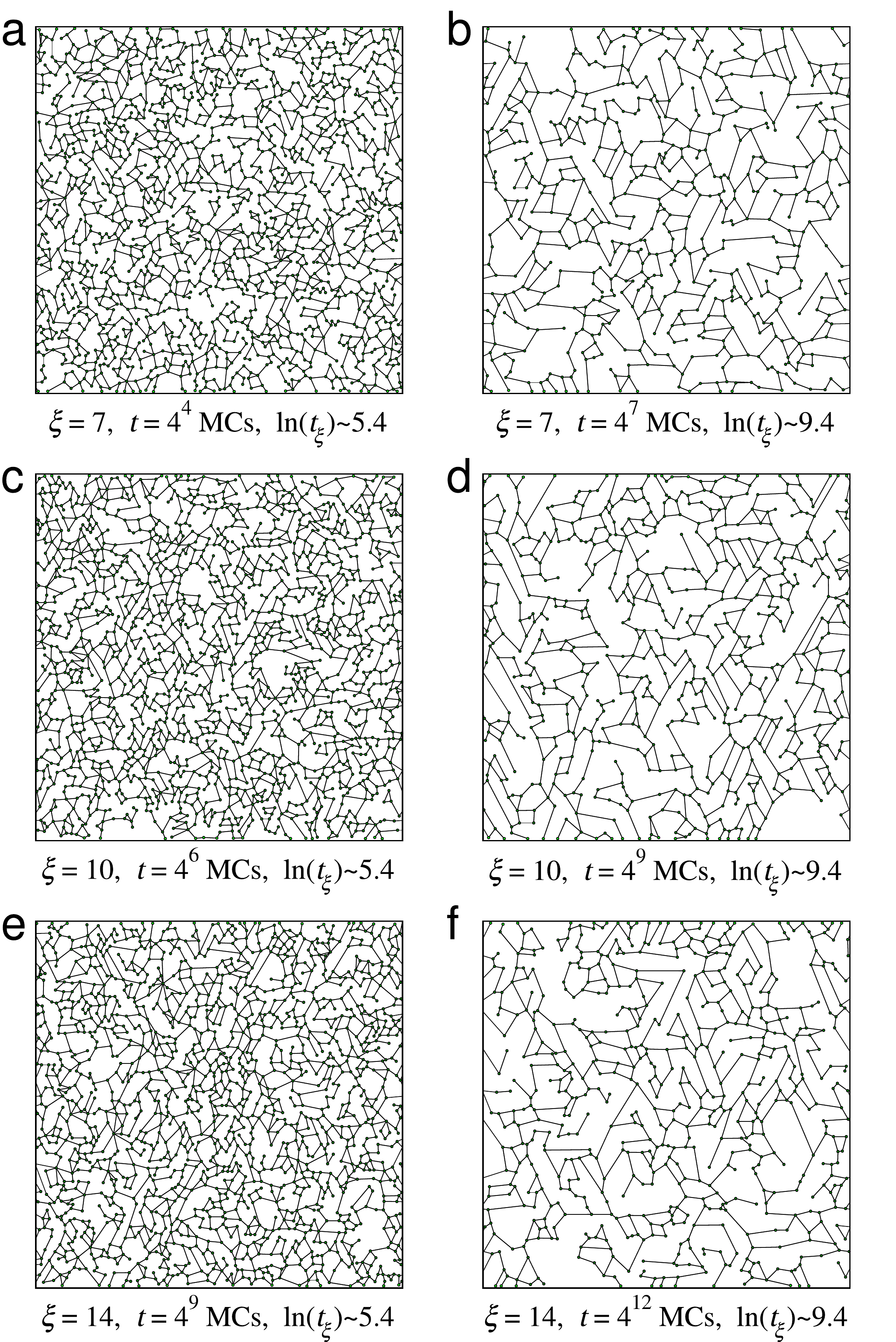}
\caption{{\bf Universal reduction in network connectivity.} Representative configurations of the amyloid fibril network at two different rescaled times. Configurations {\bf a, c,} and {\bf e} corresponds to rescaled time $\ln(t_{\xi}) \sim 5.4$ (near the maximum of the normalized shear modulus $G/E^f$ in Fig. \ref{scaling_shearGGaff}), while configurations {\bf b, d,} and {\bf f} corresponds to rescaled time $\ln(t_{\xi}) \sim 9.4$ (near the minimum of $G/E^f$ in Fig. \ref{scaling_shearGGaff}). All configurations were obtained for a coverage $\theta=0.5$.}
\label{snapshots_SI}
\end{figure}

\begin{figure}[h!]
\centering
\includegraphics[width=0.97\textwidth]{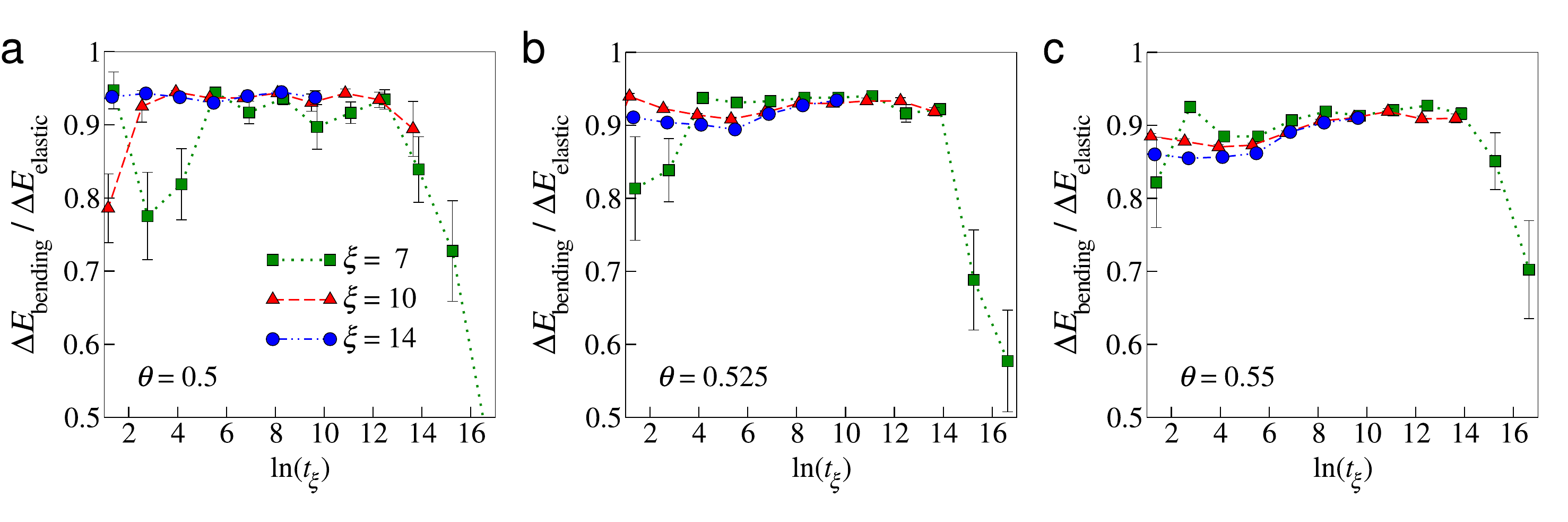}

\vspace{-0.5cm}

\caption{
{\bf Fraction of bending elastic energy.} Scaled time evolution of the ratio between the changes in the bending and total elastic energies of the network.
Panels {\bf a, b, c} display curves obtained for networks with coverages $\theta=0.5$, 0.525, and 0.55, respectively. Averages and errors bars obtained from 25 independent simulations.}
\label{energyratio}
\end{figure}

\end{document}